\documentclass[11pt]{article}
\usepackage{amssymb,cite,graphics}
\newcommand\blank[1]{}

\newcommand\toline[1]{--#1}

\newcommand{\fract}[2]{{\textstyle\frac{#1}{#2}}}
\newcommand{\fr}[2]{{\textstyle\frac{#1}{#2}}}
\newcommand{\ri}{\right}

\newcommand{\lf}{\left}
\newcommand{\te}{\theta}
\newcommand{\CS}{{\cal S}}
\newcommand\ZZ{{\mathbb Z}}

\newcommand{\CA}{{\cal A}}

\newcommand\eq{\begin{equation}}
\newcommand\en{\end{equation}}
\newcommand\bea{\begin{eqnarray}}
\newcommand\eea{\end{eqnarray}}
\newcommand\nn{\nonumber}

\newcommand{\wt}{\widetilde}
\newcommand{\FT}[1]{{\cal F}\!\left[#1\right]}
\newcommand{\w}{\omega}
\newcommand{\g}{\gamma}
\newcommand{\One}{{\hbox{{\rm 1{\hbox to 1.5pt{\hss\rm1}}}}}}
\renewcommand{\One}{{\mathbb 1}}
\renewcommand{\One}{{\rm 1\!\!1}}

\newcommand{\resection}[1]{\setcounter{equation}{0}\section{#1}}

\begin{document}
\begin{titlepage}
\vskip 0.5cm
\begin{flushright}
DTP/00/53  \\
T00/094  \\
ITFA 00-14 \\
{\tt hep-th/0008039}\\
\end{flushright}
\vskip 1.8cm
\begin{center}
{\Large {\bf Differential equations for general}} \\[5pt]
{\Large {\bf $SU(n)$ Bethe ansatz systems} }
\end{center}
\vskip 0.8cm

\centerline{Patrick Dorey%
\footnote{{\tt p.e.dorey@durham.ac.uk}},
Clare Dunning%
\footnote{{\tt t.c.dunning@durham.ac.uk}}
and Roberto Tateo%
\footnote{{\tt tateo@wins.uva.nl}}}
\vskip 0.9cm
\centerline{${}^1$\sl\small SPhT Saclay,
91191 Gif-sur-Yvette cedex, France\,}
\vskip 0.2cm
\centerline{${}^{1,2}$\sl\small Dept.~of Mathematical Sciences,
University of Durham, Durham DH1 3LE, UK\,}
\vskip 0.2cm
\centerline{${}^3$\sl\small Universiteit
van Amsterdam, Inst.~voor Theoretische
Fysica, 1018 XE Amsterdam, NL\,}
\vskip 1.25cm

\vskip 0.9cm
\begin{abstract}
\vskip0.15cm
\noindent
We show that $SU(n)$ Bethe Ansatz
equations with arbitrary `twist' parameters are hidden
inside certain $n^{\rm th}$ order ordinary differential equations,
and discuss various consequences of this fact.
\end{abstract}
\end{titlepage}
\setcounter{footnote}{0}
\def\thefootnote{\fnsymbol{footnote}}
\resection{Introduction}
Recent work has revealed unexpected connections between the functional 
relations and Bethe ansatz (BA) equations arising in integrable quantum field
theory, and the behaviour of certain ordinary differential equations 
(ODEs)~\cite{DTa,BLZa,Sa,DTb,DTc,Sb,Sc}. 
In this paper
we extend these links to cover a whole class of BA systems
associated with the $SU(n)$ Lie algebras. 
The papers just cited, and earlier work such 
as~\cite{Sha,Bax,Voros,BLZ2}, 
should be
consulted for the background to these developments, but we
begin with a brief review of the main points.

The first example to be found
concerned $SU(2)$ BA systems at a specific value of the
twist parameter, and associated them with certain Schr\"odinger 
equations~\cite{DTa}. The generalisation to arbitrary twist
turned out to require
the addition of an angular-momentum term to the ODE~\cite{BLZa}. 
In studies of the Bethe ansatz, a functional equation
found by Baxter, the T-Q relation, is known to encode the BA equations in a
particularly neat way; the place of this relation in the differential
equations side of the story was found in~\cite{DTb}. After this, the
question of generalisations to $SU(n)$ with $n>2$ was
addressed in \cite{DTc,Sb}. In \cite{DTc}, a reasonably complete mapping 
of the $SU(3)$ case onto third-order ODEs was found, and it was suggested
that higher-order equations would be the appropriate setting to search for
the more general case. Independent work by
Suzuki~\cite{Sb} provided some more concrete support for this idea.
However, this latter paper did not treat the completely general case (both
the deformation parameter
and the twists were restricted) and the role of a
particularly  important generalisation
of the T-Q relation, sometimes called the `dressed vacuum form',
was not elucidated. For this reason,
we return to the topic here, and show how the approach of \cite{DTc} can be
extended to general~$n$.
One of the main new features 
is a way of incorporating angular-momentum type terms
into the higher-order ODEs which encodes the
BA twists in a particularly neat way.
This simplifies the
discussion considerably, and even streamlines aspects of
the $SU(3)$ case treated in \cite{DTc}. We also derive nonlinear integral
equations (NLIEs) for the associated spectral problems, and discuss duality
properties.

\smallskip
\resection{The ordinary differential equation}
The initial ODE that one might consider has the form
\eq
\Bigl((-1)^{n+1}\frac{d^n}{dx^n}+P(x,E) \Bigr)\psi(x)=0~,\qquad P(x,E)=x^{nM}-E
\label{nde}
\en
with $M$ a positive real number, related to the deformation parameter. 
For $n{=}3$ this is the first of the equations
studied in~\cite{DTc}, while the general-$n$ case, but with $nM$ restricted
to integer values, was the subject of \cite{Sb}. For each $n$, (\ref{nde})
has only one free parameter, namely $M$,
and so it is clear that this equation cannot hope
to incorporate $SU(n)$ BA systems at general values of the twists.
The results of \cite{BLZa,DTc} would tend to
suggest that these twists should be associated
with the addition of terms homogeneous with $\frac{d^n}{dx^n}$, of the form
$A_kx^{-k}\frac{d^{n-k}}{dx^{n-k}}$, $k=2\dots n$. (It will be
convenient to assume that the term with $k=1$ has
been eliminated by a 
suitable redefinition of $\psi$.)
This gives $n{-}1$ further free
parameters, exactly matching the number of twists in an $SU(n)$ BA system.
However, the variables $\{A_k\}$ turn out to be rather inconvenient, and it
is better to take a slightly more indirect route.

First, define a general homogeneous differential operator of
degree one by setting
\eq
D(g)=\left(\frac{d}{dx}-\frac{g}{x}\right)~.
\en
Useful properties of this operator are
\eq
D(g)^{\dagger}=-D(-g)~,
\label{adjprop}
\en
\eq
D(g_2-1)D(g_1)=D(g_1-1)D(g_2)~.~~~~
\en
The first of these relates the operator to its adjoint, while the second 
expresses a form of commutativity. Now, given a vector ${\bf g}=(g_0,g_1,\dots
g_{n{-}1})$, set
\eq
D({\bf g})=D(g_{n-1}-(n{-}1))\,D(g_{n-2}-(n{-}2))\,\dots\,
D(g_1-1)\,D(g_0)~.
\label{dfactdef}
\en
This is a homogeneous differential operator of order $n$; by the
`commutativity' property it depends on the components
$\{g_0\dots g_{n-1}\}$ of ${\bf g}$ in a symmetrical manner. 
{}From now on we also impose
\eq
\sum_{i=0}^{n-1}g_i=\frac{n(n{-}1)}{2}
\label{van}
\en
to ensure the vanishing of the term in $D({\bf g})$
proportional to $x^{-1}\frac{d^{n-1}}{dx^{n-1}}\,$.
Finally, we record one more property of $D({\bf g})$ which will be useful
later:
its indicial polynomial (see for example~\cite{CL}) is 
\eq
f(\lambda)=\prod_{i=0}^{n-1}(\lambda-g_i)~.
\label{indic}
\en
With these ingredients in place, the $n^{\rm th}$ order ODE that we propose
to study is obtained simply by replacing the operator $\frac{d^n}{dx^n}$
in (\ref{nde}) by $D({\bf g})\,$:
\eq
\Bigl((-1)^{n+1}D({\bf g})+P(x,E) \Bigr)\psi(x)=0~,\qquad P(x,E)=x^{nM}-E~.
\label{gnde}
\en
%
%
\smallskip
\resection{The fundamental system of solutions}
As in \cite{DTb,DTc,Sb,Sc} we take our cue from
the approach of~\cite{Sha} to second-order ODEs.
We claim that (\ref{gnde}) has a solution $y(x,E,{\bf g})$ 
such that:

{\parindent=3pt
$\bullet$  $y(x,E,{\bf g})$ is an entire function of
$(x,E,{\bf g})$, modulo a possible branch point at the origin of the
complex $x$ plane.

$\bullet$  as $|x|\to\infty$ in the sector $|\arg
x|<(n{+}1)\pi/n(M{+}1)$\,,
\eq
\frac{d^py}{dx^p}\sim (-1)^p\,
\frac{\, x^{(1{-}n{+}2p)M/2}}{i^{(n{-}1)/2}_{\phantom{a}}\sqrt n}
\exp(-x^{M+1}/(M{+}1))
\label{asy}
\en
for $p=0,1,\dots$\,.
This holds if $M>1/(n{-}1)$; more generally the WKB-like
formula 
$y\sim P(x,E)^{-(n-1)/2n} \exp\bigl( - \int^{x}
P(t,E)^{1/n} dt\bigr)$ 
can be used in the way  explained,
for $n{=}2$, in the appendix of \cite{DTb}.
The normalisation of $y$ has been chosen 
for later convenience.

$\bullet$  
$y$ is uniquely characterised by
the asymptotic (\ref{asy}) for $p=0$ and $x$ real.
}

\smallskip
\noindent
For $n{=}2$, $nM{\in}\ZZ$, and $D({\bf g}){=}\frac{d^2}{dx^2}\,$, 
these properties were proved
in \cite{Sha}. We have not
attempted a proof for the
more general case; however, see \cite{Sb} for a related discussion.

\medskip
\noindent
To construct further solutions, 
we set\footnote{Note that since 
$\omega^{n(M{+}1)}=1$, the shift in $E$ in the definition of $y_k$
could equally have been 
written as $\omega^{nk}$, so long as $k$ remains an integer. This would 
have been more in line with \cite{BLZ2,DTa,DTb},
and emphasises the fact
that in the
\lq semiclassical' limit $M\to\infty$, the shifts tend to $1$.
Here we retain the conventions  of \cite{DTc}, since this makes the
transition to BA equations in standard form a little simpler, and also
appears to be the  most natural
choice when half integer values of $k$ come to be discussed, in \S7 below.}
\eq
y_k(x,E,{\bf g})=\omega^{(n{-}1)k/2} \,
y(\omega^{-k}x,\omega^{-nMk}E,{\bf g})~~,\qquad\omega=e^{2\pi i
/n(M{+}1)}\,.
\label{yk}
\en
For $k\in\ZZ$, it is easily checked that $y_k$ solves
(\ref{gnde}). If sectors  $\CS_k$ are defined as
\eq
\CS_k: \qquad |\arg x-\frac{2k\pi}{n(M{+}1)}|<\frac{\pi}{n(M{+}1)}~,
\en
then 
as $x\to\infty$ in $\CS_{k-\frac{n}{2}}\cup\CS_{k-\frac{n}{2}+1}
\cup \dots \cup \CS_{k+\frac{n}{2}}$\,, 
(\ref{asy}) implies that
\eq
\frac{d^p y_k}{dx^p} \sim (-1)^p 
\omega^{-k(1{-}n{+}2p)(M{+}1)/2}_{\phantom{a}}
\frac{\, x^{(1{-}n{+}2p)M/2}}{i^{(n{-}1)/2}_{\phantom{a}}\sqrt n}
\exp(-\omega^{-k(M{+}1)}x^{M{+}1}/(M{+}1))~.
\label{ykasymp}
\en
We call {\it subdominant} a
solution which tends to zero fastest in a given sector; up to a constant
multiple, $y_k$ is the unique solution to (\ref{gnde})
which is subdominant in $\CS_k$.

Wronskians between these solutions will be important in the
following, so we define
\eq
W^{(m)}_{k_1,k_2,\dots k_m} = W^{(m)}[y_{k_1},y_{k_2},\dots y_{k_m}]\,,
\label{wmdef}
\en
where 
\eq
W^{(m)}[f_1,f_2,\dots f_m]=
\mbox{Det}
\lf[\matrix{f_1 & f_2 & \dots& f_{m}  \cr
f_1' & f_2' & \dots& f_{m}'  \cr
\vdots & \vdots &  & \vdots  \cr
f_1^{[m-1]}  &  f_2^{[m-1]} & \dots &
f_m^{[m-1]}   }\ri]
\en
and $f_i^{[p]}(x)=\frac{d^p}{dx^p}f_i(x)$.
We will give a special status to those Wronskians whose arguments are
successive integers, and in preparation for this we set
\eq
W^{(m)}_k=W^{(m)}_{k,k+1,k+2,\dots\, k+m-1}~.
\label{spec}
\en
Note that $W^{(1)}_k=y_k$, and that (generalising (\ref{yk}))
\eq
W^{(m)}_{k_1+k , k_2+k , \dots \ k_m+k}(x,E,{\bf g})=\omega^{m(n{-}m)k/2}
W^{(m)}_{k_1 , k_2 , \dots \ k_m}(\omega^{-k}x,\omega^{-nMk}E,{\bf g})~.
\label{pwron}
\en

For an $n^{\rm th}$ order ODE with vanishing $(n{-}1)^{\rm th}$ order
term, as is true of our case by~(\ref{van}), it is standard that all
$n$-fold Wronskians $W^{(n)}[f_1,\dots f_n]$ are 
independent of $x$, and vanish if and only if the solutions $f_1,\dots
f_n$ are linearly dependent.
In particular, (\ref{ykasymp}) can be used as $|x|\to\infty$ 
in $\CS_{k-1/2}\cup\CS_{k+1/2}$ to show that
$W^{(n)}_k=1$, and so, for each $k\in\ZZ$, the functions
$\{y_k , y_{k+1} ,\dots ,y_{k+n-1}\}$ are linearly independent, and 
furnish a basis of solutions to (\ref{gnde}). 
%
%
\smallskip
\resection{Stokes multipliers and functional relations}
We next establish a basic functional relation. 
Expanding $y_0$ in the basis provided by
$\{y_1 , y_2 ,\dots ,y_n\}$,
it must be possible to write
\eq
\sum_{k=0}^{n} (-1)^k C^{(k)}(E,{\bf g})\, y_k(x,E) = 0  ~,
\label{nstokes}
\en
where 
$C^{(0)}(E,{\bf g})=1$ and
the Stokes multipliers $C^{(k)}(E,{\bf g})$, $k>0$
are analytic functions of $E$ and ${\bf g}$. (The formal similarity
between this equation and equation (5.50) of \cite{KLWZ} should be noted.)
Given that
$W^{(n)}_{1}=1$,
we have the simple relation
\eq
C^{(k)}(E,{\bf g})= 
W^{(n)}_{0 1 \dots \hat k \dots n} (E,{\bf g})
\en
with the hat ($\hat k$) indicating that the corresponding index is
to be omitted. Clearly, $C^{(n)}=W^{(n)}_0=1$. 
To treat the other Stokes multipliers,
we first relate them to
the `privileged' Wronskians $W^{(m)}_k$. We will only need
$C^{(1)}$ here, but the discussion can be generalised to the other cases.
For general $p$, consider the determinants
\eq
0=
\mbox{Det}
\lf[\matrix{
y_0^{[i]}  & y_1^{[i]}  & \dots &  y_{p}^{[i]}  \cr
y_0  & y_1  & \dots &  y_{p}  \cr
y_0' & y_1' & \dots &  y'_{p}  \cr
\vdots & \vdots   &       &  \vdots \cr
y_0^{[p-1]} & y_1^{[p-1]} & \dots &
y_{p}^{[p-1]}  }\ri]
\en
with $i=0,1,\dots p{-}2$. 
Expanding the first row of each using Cramer's rule, we have
\eq
0{~}
={~}
\sum_{k=0}^{p}\,
(-1)^{k} 
\lf[\matrix{y_k\cr y_k'\cr\vdots\cr y_k^{[p-2]}}\ri]
W^{(p)}_{01\dots \hat k\dots p-1} ~.
\label{cram}
\en
Now form a $(p{-}1){\times}(p{-}1)$ matrix 
with the RHS of (\ref{cram}) as the first column, and the
vectors ${\bf v}_j=[y_j,y_j',\dots y_j^{[p-2]}]^t$, $j=2,3\dots p{-}1$ 
as the remainder. Taking its determinant yields the following Plucker-type
relation:
\eq
0=
W^{(p-1)}_{02\dots p-1} W^{(p)}_{12 \dots p} -
W^{(p-1)}_{12\dots p-1} W^{(p)}_{02 \dots p} + (-1)^{p}
W^{(p-1)}_{p2\dots p-1} W^{(p)}_{01 \dots p-1} ~.
\en
Rearranging, and supplementing
the notation (\ref{spec}) with the convention
$W^{(0)}_k=1$ $\forall k\,$,
\eq
\frac{W^{(p)}_{02\dots p}}{W^{(p)}_1}
{~} = {~} 
\frac{W^{(p-1)}_{02\dots p-1}}{W^{(p-1)}_1}
\,+\,
\frac{W^{(p-1)}_2}{W^{(p-1)}_1}\,
\frac{W^{(p)}_0}{W^{(p)}_1}
{~} = {~} 
\sum_{m=0}^{p-1}
\frac{W^{(m)}_2}{W^{(m)}_1}\,
\frac{W^{(m+1)}_0}{W^{(m+1)}_1}~,
\label{interm}
\en
the second equality being obtained by repeated substitution.
This is the result we need. 
Since
$C^{(1)}=W^{(n)}_{02\dots n}\,$, we can
set $p=n$ and
multiply (\ref{interm})
through by $\prod_{j=0}^{n}W^{(j)}_1$ to find
\eq
C^{(1)}\prod_{j=0}^{n} W^{(j)}_1 
{~} ={~}
\sum_{m=0}^{n-1}~ 
\Biggl(
\prod_{j=0}^{m-1} W^{(j)}_1
\Biggr)
W^{(m)}_2
W^{(m+1)}_0 
\Biggl(
\prod_{j=m+2}^{n}\! W^{(j)}_1
\Biggr)~.
\label{t1reln}
\en
%
%
\smallskip
\resection{Bethe ansatz equations}
 As it stands, equation (\ref{t1reln})
involves $x$ as well as $E$ and ${\bf g}$, and so is not quite ready to be
mapped onto a standard set of BA equations. 
The simplest idea, simply to set $x$ to zero, only works at exceptional
values of the $g_i\,$, since in general
the differential operator
$D({\bf g})$ 
has a (regular) singularity at $x=0$.
Instead, motivated by the treatments of 
\cite{BLZa,DTb,DTc}, we expand 
$y(x,E,{\bf g})$ as
\eq
y(x,E,{\bf g})=
W^{(1)}_0(x,E,{\bf g})= \sum_{i=0}^{n-1}  
D^{(1)}_{[i]}(E,{\bf g})\,\chi_i(x,E,{\bf g})
\label{yexp}
\en
where the $\chi_i$ form an alternative
basis of solutions to (\ref{nde}), fixed by the demand that they have
the simplest possible behaviours near the origin:
\eq
\chi_i \sim x^{g_i} + O(x^{g_i+n})\,,
\qquad x\to 0\,.
\label{chas}
\en
(Recall that the $g_i$ are
the roots of the indicial polynomial (\ref{indic}).)
Strictly speaking the asymptotic (\ref{chas}) does not always suffice
to pin down {\em all} of the $\chi_i$. 
Assume, until further notice, that
the $g_i$'s are ordered as
\eq
\Re e (g_0) < \Re e ( g_1) < \dots <\Re e (g_{n-1})\,.
\label{gord}
\en
Then $\chi_0$ is certainly uniquely determined by
(\ref{chas}), and the $\chi^{\phantom a}_{i>0}$ can
be defined by a process of analytic continuation from this solution,
just as 
is done for the radially-symmetric Schr\"odinger equation
when passing between the so-called regular and irregular solutions.
This will be used in \S8 below.

The direct substitution of (\ref{yexp}) into (\ref{wmdef})
yields an expansion for $W^{(m)}_0(x,E,{\bf g})$ in which the functions
$\chi_{i}(\w^{-k}x,\w^{-nMk}E,{\bf g})$ appear
as well as the
$\chi_{i}(x,E,{\bf g})$. However, all of these functions are solutions
to the initial ODE, and by considering their behaviour near $x=0$ one
finds
\eq
\chi_{i}(\w^{-k}x,\w^{-nMk}E,{\bf g})
=
\w^{-kg_i}\chi_{i}(x,E,{\bf g})\,.
\label{chiprop}
\en
Therefore the $m$-fold Wronskians $W^{(m)}_0$ have expansions of the form
\eq
W^{(m)}_0(x,E,{\bf g})=
\sum_{0\le j_1<j_2<\dots<j_m\le n{-}1}
D^{(m)}_{[j_1 j_2 \dots j_m]}(E,{\bf g}) 
\,W^{(m)}[\chi_{j_1},\chi_{j_2}\dots \chi_{j_m}](x,E,{\bf g})~.
\label{wexp}
\en
Using (\ref{chiprop}), (\ref{yexp}) and (\ref{wmdef}), the coefficients
$D^{(m)}_{[j_1 j_2 \dots j_m]}(E,{\bf g})$ can be expressed as sums of
products of the $D^{(1)}_{[i]}(\w^{-nMk}E,{\bf g})$. This leads to relations
which generalise the `quantum Wronskians' of \cite{BLZ2}. However, for
current purposes it is better to treat the coefficients with different
values of $m$ as independent functions, and so we will leave further 
discussion
of this point to another occasion.

We will initially focus on the 
dominant terms of the expansions (\ref{wexp}). With the ordering 
(\ref{gord}), these are
\eq
W^{(m)}_0(x,E,{\bf g})\sim
D^{(m)}_{[01\dots m-1]} (E,{\bf g})
\, x^{\beta_m+m(n-m)/2}\,,
\qquad x\to 0\,
\en
where in order to simplify subsequent calculations we set
\eq
\beta_m=\sum_{j=0}^{m-1} g_j
-m(n{-}1)/2~.
\en
Substituting (\ref{wexp}) into (\ref{t1reln}),
an $x$-independent equation is found
by extracting the coefficient 
of the leading power 
$x^{\alpha}$, $\alpha =  \sum_{j=0}^{n}
( \beta_j +j(n{-}j)/2 )\,$.
Shifting
$E$ to 
$\w^{nM}E$ and setting
\bea
&& D^{(m)}(E,{\bf g})\,=\, D^{(m)}_{[01\dots m-1]} (E,{\bf g})
{}~,\quad
 D^{(m)}_k(E,{\bf g})\,=\, 
D^{(m)}(\w^{-nMk}E,{\bf g})\,,\nn\\[3pt]
&& T^{(1)}(E,{\bf g})~\,\,=\, 
C^{(1)}(\w^{nM}E,{\bf g})~,
\label{TCDkD}
\eea
the final result can be written as
\eq
T^{(1)}\prod_{j=0}^{n} D^{(j)}_0
{~}={~}
\sum_{m=0}^{n-1}\, 
\Biggl(
\prod_{j=0}^{m-1} D^{(j)}_0
\Biggr)\,
\w^{-\beta_m}D^{(m)}_{1}\,
\w^{\beta_{m+1}}D^{(m+1)}_{-1}
\Biggl(
\prod_{j=m+2}^{n}\! D^{(j)}_0
\Biggr)~.
\label{cdsys}
\en
This is a generalised T-Q relation, with
the $D^{(j)}$ playing the role of the eigenvalues of
the $Q$ (or
$A$ in \cite{BLZ2})
operators.  In the language of integrable lattice models relations of
this type are known as the 
dressed vacuum form for the transfer matrix eigenvalues (see for example 
\cite{Res,KStq}).

We can now derive an initial set of BA equations.
Suppose that the zeroes of $D^{(m)}(E)$ are at $E=F^{(m)}_k$,
$k=1,2\dots\,\infty$. (For the next few equations we will leave the dependence
of all functions on ${\bf g}$ implicit.) Setting $E=F^{(m)}_k$ 
in (\ref{cdsys}), the LHS vanishes, as do all but two terms in the sum
on the RHS. Assuming, as will generically be the case, that there are
no further vanishings,
this gives us the following set of coupled equations
for the $\{F^{(m)}_k\}\,$:
\eq
\frac{D^{(m-1)}(\w^{-nM} F^{(m)}_k)}
     {D^{(m-1)}(F^{(m)}_k)~}
\,
\frac{D^{(m)}(\w^{nM}F^{(m)}_k)}
     {D^{(m)}(\w^{-nM} F^{(m)}_k)}
\,
\frac{D^{(m+1)}(F^{(m)}_k)~}
{D^{(m+1)}(\w^{nM}F^{(m)}_k)}
\,
=
\,
-\w^{-2\beta_m+\beta_{m-1}+\beta_{m+1}}_{\phantom l}\,.
\label{baea}
\en
These form a system of Bethe Ansatz equations (BAE)
of $SU(n)$ type (see, for example,
equation (3.32) of \cite{KLWZ}).
However the reality properties are not
very transparent when the equations are given 
in this form, and it is useful to make one more
definition, setting
\eq
D^{(m)}(E,{\bf g})=
A^{(m)}( \w^{- nM(m-1)/2} E,{\bf g})~,\quad
F^{(m)}_k = \w^{ nM(m-1)/2} E^{(m)}_k \,.
\label{adefn}
\en
Then the BAE (\ref{baea}) become
\eq
\frac{A^{(m-1)}(\w^{-nM/2}E^{(m)}_k)}
     {A^{(m-1)}(\w^{nM/2}E^{(m)}_k)~}
\,
\frac{A^{(m)}(\w^{nM}E^{(m)}_k)}
     {A^{(m)}(\w^{-nM}E^{(m)}_k)}
\,
\frac{A^{(m+1)}(\w^{-nM/2}E^{(m)}_k)~}
     {A^{(m+1)}(\w^{nM/2}E^{(m)}_k)}
\,
=
\,
-\w^{-2\beta_m+\beta_{m-1}+\beta_{m+1}}_{\phantom l}.
\label{baeb}
\en
These can also be
written using the Cartan
matrix $C_{mt}$ of the $SU(n)$ Dynkin diagram:
\eq
\prod_{ t=1}^{n-1} 
\omega^{C_{mt}\beta_t}_{\phantom a}
\frac
{A^{(t)}(\omega^{\frac{nM}{2} C_{mt}}E^{(m)}_{k})}
{A^{(t)}(\omega^{-\frac{nM}{2}C_{mt} }E^{(m)}_{k})}= 
-1\,,\qquad k=1,2,\dots~.
\label{dall}
\en
Finally, the BAE can be given a factorised form once the growth
properties of the functions involved have been established.
{}A WKB treatment along the lines of \cite{DTb} shows that
the function $A^{(1)}(E,{\bf g})=D^{(1)}_{[0]}(E,{\bf g})$ has the 
large $|E|$ asymptotic
\eq
\ln A^{(1)}(E,{\bf g}) 
\sim  \kappa(nM,n) (-E)^{\mu}, \qquad |E| 
\to \infty\,,~ |\arg(-E)| < \pi~,
\label{asa1}
\en
where 
\eq
\mu={(M{+}1) \over nM} \, \, , \ \ \ \kappa(a,b)=
\frac { \Gamma(1+\frac{1}{a}) \Gamma(1+\frac{1}{b}) \sin(\frac{\pi}{b}) }
{ \Gamma(1+\frac{1}{a}+\frac{1}{b})  \sin(\frac{\pi}{b}+\frac{\pi}{a}) }~.
\en
The asymptotics for the remaining  $\ln A^{(m)}(E,{\bf g})$ are more 
tricky. So long as it is assumed that the other functions $D^{(1)}_{[i]}$
share the asymptotic (\ref{asa1}), then the quantum 
Wronskian-like relations mentioned just after (\ref{wexp}) can be used to
show that all of the  $A^{(m)}(E,{\bf g})$ are of the same order, namely $\mu$.
For this section this suffices, but for the NLIE a more precise result will be
required. Unfortunately, cancellations of leading terms
seem to be at work inside the quantum
Wronskians in almost all directions in the complex
$E$ plane, making a direct calculation rather difficult.
However, indirect evidence suggests  the following:
\eq
\ln A^{(m)}(E, {\bf g}) \sim   
\frac{\sin(\pi m/n)}{\sin(\pi/n)} \kappa(nM,n) (-E)^{\mu}~~,\quad|\arg(-E)| < \pi~.
\label{gas}
\en
Note that this claim matches the $\ZZ_2$  symmetry property 
of the spectral determinants discussed in  \S 7 below,
and  its  implications are in agreement with the numerical checks for the 
``soluble''  $M=1/n$ cases reported in \S8.

For $M>1/(n{-}1)$ the order $\mu$ of
the functions $A^{(m)}(E,{\bf g})$
is less than one, so 
the Hadamard factorization theorem implies
\eq
A^{(m)}(E,{\bf g} )=A^{(m)}(0,{\bf g})
 \prod_{j=1}^\infty \Bigl(1-\frac{E}{E^{(m)}_j}\Bigr) ~.
\label{nhad}
\en
This finally allows (\ref{dall}) to be written
as
\eq
\prod _{t=1}^{n-1} 
\omega^{C_{mt}\beta_t}
\prod_{j=1}^\infty
\left(
\frac{ E^{(t)}_{j} -\omega^{\frac{nM}{2} C_{mt}}E^{(m)}_k}
{ E^{(t)}_{j} -\omega^{-\frac{nM}{2} C_{mt}}E^{(m)}_k} 
\right)= 
-1\,,\qquad k=1,2,\dots~.
\label{bae}
\en
\smallskip
\resection{The nonlinear integral equation}
Next we derive a set of coupled nonlinear integral equations. Define
\eq
a^{(m)} (E,{\bf g})=
\prod_{t=1}^{n-1} 
\omega^{-C_{mt}\beta_t}
\frac{A^{(t)} (\omega^{-\frac{nM}{2} C_{mt}}E,{\bf g})}
{A^{(t)} (\omega^{\frac{nM}{2} C_{mt}}E,{\bf g})}
\en
so that $a^{(m)} (E^{(m)}_k\!,{\bf g})=-1$ by  
(\ref{dall}).
We will follow the ideas of \cite{DDV,BLZ2} (another approach
can be found in \cite{KBP}). The product representation (\ref{nhad})
allows
$\ln a^{(m)}$ to be written as an infinite sum over the zeroes $E^{(m)}_k$
of $A^{(m)}$. We now make two important assumptions: that all of the
$E^{(m)}_k$
lie on the positive real axis of the complex $E$ plane, and that these are the 
only points in some narrow strip about this axis at which $a^{(m)}(E,{\bf g})$ is
equal to $-1$.
We expect that these will hold for some range of the parameters ${\bf g}$.
Cauchy's theorem can then be used to replace
the sum by an integral
along a contour $\cal C$ which runs from $+\infty$ 
to $0$ above the real axis, encircles the origin and 
returns to $\infty$  below the real axis:
\eq
\ln a^{(m)}(E,{\bf g})=
\frac{-\,2\pi i~}{n(M{+}1)}  
\, \sum_{t=1}^{n-1}
C_{mt}\,\beta_{t}
\,
+ \sum_{t=1}^{n-1} 
\int_{\cal C} \frac{dE'}{2\pi i} F_{mt}(E/E')
\partial_{E'}\ln(1+a^{(t)}(E',{\bf g})) 
\en
where 
\eq
F_{mt}(E)=\ln \frac{ 1-\omega^{-\fract{nM}{2} C_{mt}}E}{1-\omega^{\fract{nM}{2} C_{mt}}E}~~. 
\en
After a variable change 
$E=\exp(\theta/\mu)\,$,
we define (with a mild abuse of notation)
$\ln a^{(m)}(\theta)  
\equiv \ln a^{(m)}(e^{\theta/\mu},{\bf g})\,$,
integrate by parts and use the property  $[a^{(m)}(\theta)]^*
=a^{(m)}(\theta^*)^{-1}$ to find
\bea
\ln a^{(m)}(\theta)
-\sum_{t=1}^{n-1} \int_{-\infty}^{\infty} d\theta' R_{mt}(\theta-\theta') 
\ln a^{(t)}(\theta')&=& 
\frac{-2\pi i}{n(M{+}1)} 
\,
\sum_{t=1}^{n-1}C_{mt}\,\beta_{t}
 \\ \nn
&&\hskip-3.5cm {}-2i \sum_{t=1}^{n-1}
\int_{-\infty}^{\infty} d\theta' R_{mt}(\theta{-}\theta')
\Im m \ln (1+a^{(t)}(\theta'{-}i 0))~.  
\label{eq}
\eea
Here $R_{mt}(\theta)=
\frac{i}{2\pi}\partial_{\theta}F_{mt}(e^{\theta/\mu})$.
We now take the Fourier transform of this equation,
setting
\bea
\wt f (k) &=&
\FT f (k)  \ \ \ \,\,= \ \int_{-\infty} ^{\infty} f(\te) \, 
e^{-i\,\te\,k} ~ d\te~,   \\
f(\te) &=&
{\cal F}^{-1} [ \wt{f}]\,(\theta)\ = \
\frac{1}{2\pi} \, 
\int_{-\infty} ^{\infty} \wt{f}(k)%
\, e^{i \, \te \, k} ~ dk ~.
\eea
The transformed equations can be written
in a compact form as 
\bea
\sum_{t=1}^{n-1} 
(\delta_{mt}-\wt{R}_{mt}(k)) \FT{\ln\,a^{(t)}}\!(k) &=& 
{}- \frac{ 4 i \pi^2 \delta(k)}{n(M{+}1)} 
\,
\sum_{t=1}^{n-1} C_{mt}\,\beta_{t} 
\nn \\
&&{~~}- 2i \sum_{t=1}^{n-1}
\wt{R}_{mt}(k)
 \Im m \FT{\ln\,(1+ a^{(t)})}\!(k)~, \qquad
\eea
where the non-vanishing entries of $\wt{R}_{mt}(k)$ are
\eq
\wt{R}_{<mt>}(k) = 
{\sinh( \fract{\pi}{n} \xi k) \over \sinh( \fract{\pi}{n} (1+ \xi) k)}
~~,~~
\wt{R}_{mm}(k) ={\sinh( \fract{\pi}{n} (1- \xi) k) \over 
\sinh( \fract{\pi}{n} (1+ \xi) k)}~~,\quad (\,\xi=\frac{1}{M}\,)
\label{nvani}
\en
the notation $ <mt>$ implying that the nodes $m$ and $t$ are connected 
on the Dynkin diagram of $SU(n)$.
Applying the inverse matrix $(\One-\wt R(k))^{-1}$,
transforming back and rewriting the imaginary parts in terms
of values above and below the real axis, we obtain
a set of coupled NLIEs for the functions
$a^{(m)}(\theta)$, $m=1,\dots,n{-}1$\,, considered along 
contours 
${\cal C}_1$ and ${\cal C}_2$ which run from $-\infty$ 
to $+\infty$, just below and just above the real $\theta$-axis:
\bea
&&
\hskip -25pt
\ln a^{(m)}(\theta)= 
i \pi \alpha_m({\bf g})- i b_0 M_m\, e^{\theta} \nn \\[3pt]
&&\hskip -10pt {}+\sum_{t=1}^{n-1} \left( \int_{ {\cal C}_1} d \theta'
\varphi_{mt}(\theta{-}\theta') \ln(1+a^{(t)}(\theta')) -
\int_{ {\cal C}_2} d \theta'
\varphi_{mt}(\theta{-}\theta') 
\ln(1+{1 \over a^{(t)}(\theta')}) \right).\qquad
\label{nlie}
\eea
The driving terms $i b_0 M_n e^{\theta}$ arise from zero modes,
which can be traced to the poles of
$(\One- \wt R(k))^{-1}$ at $k=i$, and their magnitudes
\eq
b_0= 2 \sin( \pi \mu)  \kappa(nM,n)~, 
\qquad M_m= {\sin( \pi m/n) \over \sin( \pi/n)}~
\en
are fixed by imposing the asymptotic (\ref{gas}).
The kernel and twist factors are, respectively,
\bea
\varphi_{mt}(\theta)&=&  {\cal F}^{-1} 
\left[ \left( \One - (\One- \wt R(k))^{-1} \right)_{mt} \right](\theta)~,
\label{varphi}\\
\alpha_{r}({\bf g})&=&-\frac{2\pi}{n(M{+}1)}
\sum_{t,m=1}^{n-1} 
\left(\One- \wt R(0)\right )_{rm}^{-1}
C_{mt}\,\beta_{t}~. 
\, 
\label{alpha}
\eea
These can be written more explicitly with the help of
the deformed Cartan matrix
\eq
C_{mt}(k) =  \left\{ \begin{array}{ll}
     2  & ~~m=t , \\
    \frac{-1~}{\cosh( \frac{ \pi k}{n})} &  ~~<mt>  \end{array} \right.
\en
and its inverse
\eq
C_{tm}^{-1}(k) = 
C_{mt}^{-1}(k) =
\frac{\coth(\frac{\pi}{n} k) \sinh(\frac{\pi}{n} (n{-}m) k)
\sinh(\frac{\pi}{n}t \, k )}{\sinh( \pi k)}~~~~~(m \ge t) ~.
\en
By (\ref{nvani}), the non-vanishing  off-diagonal  and diagonal
elements  of $\wt{R}(k)$
are related by 
$2\wt{R}_{<mt>}(k)=(1{-}\wt{R}_{mm}(k))/ \cosh( \frac{\pi k}{n})$,
and so
\eq
\delta_{mt}- \wt R_{mt}(k) = {1 \over 2} C_{mt}(k) 
(1- \wt{R}_{mm}(k))~,
\label{oner1}
\en
where
\eq
1-\wt{R}_{mm}(k) 
={2 \sinh(\fr{\pi}{n} \xi k ) \cosh(\fr{\pi}{n} k)
\over  \sinh( \fr{\pi}{n} (1+\xi) k)}~.
\label{oner2}
\en
Plugging relations (\ref{oner1}) and (\ref{oner2}) into
(\ref{alpha}) and (\ref{varphi}) we obtain
\bea
\varphi_{mt}(\theta)&=& \int_{-\infty}^\infty {dk \over 2 \pi} \
e^{i k \theta}\Bigl (\delta_{mt} - 
\frac{ \sinh(\fr{\pi}{n} (1+\xi)k)}{\sinh(\fr{\pi}{n} \xi k) 
\cosh(\fr{\pi}{n} k)}
C_{mt}^{-1}(k) \Bigr)~,
\nn \\
\alpha_m({\bf g}) &=& -\frac{2}{n}
\,
\beta_m
\,,
\eea
with $\xi=1/M$.
The final NLIEs exactly match the massless versions of
the equations found in \cite{Ma,ZJ}
in a completely different context.

\smallskip
\resection{Symmetry properties}
\noindent
The $SU(n)$ Dynkin diagram has an evident $\ZZ_2$ symmetry, swapping
nodes $m$ and $n{-}m$. This symmetry is also respected by the BAE 
so long as the
twists are transformed at the same time. We would expect to see signs
of this symmetry in the 
differential equation, but quite how is not 
immediately obvious. For example,
the first node of the Dynkin diagram is associated with the basic
solution $y(x,E,{\bf g})$ while
its partner, the $(n{-}1)^{\rm th}$ node, is associated with an 
$(n{-}1)$-fold Wronskian $W^{(n{-}1)}_0(x,E,{\bf g})$.
The answer to this puzzle is provided by the fact that taking
the Wronskian 
of $n{-}1$ solutions to an $n^{\rm th}$ order ODE automatically provides
a solution to the adjoint equation. For $n{=}3$ this observation can be 
found in \cite{Birk}, and was related to the current context in \cite{DTc}.
For general $n$, the result can be established as follows.
It is easiest not to use the factorised form (\ref{dfactdef})
of $D({\bf g})$ for the differential operator, 
but rather to suppose that (\ref{gnde}) has been rewritten as
\eq
\frac{d^n}{dx^n}\,\psi+
B_2(x)
\frac{d^{n-2}}{dx^{n-2}}\,\psi+
B_3(x)
\frac{d^{n-3}}{dx^{n-3}}\,\psi+\dots
+(-1)^{n+1}P(x,E)\,\psi\,=\,0\,.
\label{fullode}
\en
(For the differential operators discussed in \S2,
$B_k(x)=(-1)^{n+1}A_kx^{-k}$, but neither this nor the explicit
relation between the $A_k$'s and the $g_i$'s will be needed for the argument.)
Now consider $n{-}1$ solutions to this equation, 
gathered together into a vector as ${\bf f}=(f_1,f_2,\dots f_{n-1})$.
We denote by $V[a_1,a_2,\dots a_{n-1}]$ the determinant of the matrix whose
rows are the derivatives
${\bf f}^{[a_1]}, {\bf f}^{[a_1]},\dots {\bf f}^{[a_{n-1}]}$, so that
the Wronskian of $f_1,f_2\dots f_{n-1}$ is
\eq
W^{(n-1)}[f_1,f_2,\dots f_{n-1}](x,E,{\bf g})= 
V[0,1,\dots,n{-}2] \,.
\en
Differentiating once, and using the general property that 
the differential of a determinant is equal to a sum of 
determinants in which each row has been differentiated,
\eq
\frac{d}{dx}\,
V[a_1,a_2,\dots a_{n-1}]
=
\sum_{k=1}^{n-1}
V[a_1,\dots\,a_{k-1},a_k{+}1,a_{k+1},\dots\, a_{n-1}]\,,
\en
we have
\eq
\frac{d}{dx}\,
W^{(n-1)}=
V[0,1,\dots\,n{-}4,n{-}3,n{-}1]\,,
\en
all other terms vanishing by antisymmetry.
Differentiating again,
\eq
\frac{d^2}{dx^2}\,
W^{(n-1)}=
V[0,1,\dots\,n{-}4,n{-}2,n{-}1]
+
V[0,1,\dots\,n{-}4,n{-}3,n]\,.
\en
The last term can be rewritten using 
(\ref{fullode}) to substitute for ${\bf f}^{[n]}$:
\eq
V[0,1,\dots\,n{-}4,n{-}3,n]=-B_2(x)V[0,1,\dots\,n{-}4,n{-}3,n{-}2]
=-B_2(x)W^{(n-1)}.
\en
(Again, all other terms
vanish by antisymmetry.) Thus
\eq
\Bigl(\frac{d^2}{dx^2}+B_2(x)\Bigr)\,
W^{(n-1)}=
V[0,1,\dots\,n{-}4,n{-}2,n{-}1]\,.
\label{lts}
\en
The pattern should now be clear. Differentiating both sides of (\ref{lts})
will again result in two terms on the RHS. The second of these can 
be rewritten using (\ref{fullode}), but this time yields
$B_3(x)W^{(n-1)}$ instead of $-B_2(x)W^{(n-1)}$ as the one non-vanishing
contribution. Taking this over onto the LHS and continuing to differentiate,
one finally finds that $W^{(n-1)}$ satisfies
\eq
\left(\frac{d^n}{dx^n}+
\frac{d^{n-2}}{dx^{n-2}}
B_2(x)
-
\frac{d^{n-3}}{dx^{n-3}}
B_3(x)
+\dots
-P(x,E)\right) W^{(n-1)}\,=\,0\, ,
\label{fineq}
\en
which is indeed (up to an overall sign) the equation adjoint 
to (\ref{fullode}).
Now we can return to the factorised form of $D({\bf g})$,
take its adjoint using (\ref{adjprop}), and rewrite (\ref{fineq})  as:
\eq
\Bigl((-1)^{n+1}D({\bf g}^{\dagger})+ (-1)^n P(x,E) \Bigr)
W^{(n-1)}(x,E,{\bf g})=0~,
\quad P(x,E)=x^{nM}-E~,
\label{agnde}
\en
with 
\eq
{\bf g}^{\dagger}=(g^{\dagger}_0,g^{\dagger}_1,\dots\,g^{\dagger}_{n-1})~,
\qquad
g^{\dagger}_j=n{-}1-g^{\phantom{\dagger}}_{n-1-j}~.
\en
For $n$ even this is the original ODE (\ref{gnde}), modulo the swap 
to the `conjugate' set of parameters ${\bf g}^{\dagger}$. For $n$ odd, 
the term $P(x,E)$ has been replaced by $-P(x,E)$. 
However, the conventions adopted in the
definition (\ref{yk}) mean that
the `half-shifted' functions $y_{k+1/2}(x,E,{\bf g})$ ($k\in\ZZ$)
themselves solve (\ref{gnde})
with $P$ replaced by $-P$.
Taking an $(n{-}1)$-fold Wronskian of these functions
and repeating the above discussion, the minus signs cancel.
In particular, this means that whether $n$ is even or odd we have
\eq
\Bigl((-1)^{n+1}D({\bf g}^{\dagger})+ P(x,E) \Bigr)
W^{(n-1)}_{1-n/2}(x,E,{\bf g})=0~.
\label{agnde1}
\en
This is exactly the equation solved by the $y_k(x,E,{\bf g}^{\dagger})$. 
Comparing asymptotics
as $|x|\to\infty$ in ${\cal S}_0$, we finally establish the formula
\eq
W^{(n-1)}_{1-n/2}(x,E,{\bf g})=y(x,E,{\bf g}^{\dagger})\,.
\en
This shows that the $(n{-}1)$-fold Wronskian is (up to a shift in its arguments)
just a basic solution to another ODE, and thus demonstrates that
the relation between
the first and last nodes is indeed reflected in the properties of
the differential equation.
The spectral functions corresponding to the remaining nodes
on the Dynkin diagram
can now be obtained either as $m$-fold Wronskians of the $W^{(n-1)}_{k+1/2}$, or as
$(n{-}m)$-fold Wronskians of the $y_k$, and
the diagram symmetry
should be reflected in the equality of the results of the two calculations.
Keeping the normalisation of the $y$'s as in (\ref{ykasymp}), 
the necessary identity is the following:
\eq
W^{(n-m)}_{m-1}
=
W^{(m)}[W^{(n-1)}_0,W^{(n-1)}_1,\dots,W^{(n-1)}_{m-1}] \, .
\label{wida}
\en
While we do not have a general proof,
for $m=2$ this result follows
from the  Jacobi  
identity~(cf.~for example equation (2.20) of \cite{KLWZ}) 
and the fact that $W^{(n)}=1$. This case is also equivalent to a formula due to
Frobenius \cite{Frob}, more recently reviewed in \cite{Boc}.
\smallskip
\resection{Analytic continuation and the linear potential}
\noindent
Thus far, we have restricted attention to the dominant terms in 
the expansions
(\ref{wexp}). However, quite which term is dominant
depends on the ordering (\ref{gord}) of the $g_i$'s. 
While the ODE (\ref{gnde})
is insensitive
to this ordering (as remarked earlier, it depends on the $g_i$ 
in a symmetrical
manner), the same is not true of the NLIE (\ref{nlie}),
since it sees not the parameters ${\bf
g}$ but rather the BA twists ${\bf \alpha}({\bf g})$. By analytic 
continuation,
it should be possible to move between different twists corresponding 
to the same
ODE,  thereby accessing the
other terms in the expansions. In this section we will test this idea 
by means of
a simple example. This also serves as a rather strong check on the various
conjectures that have been made above.

In the expansion of $y$ given by (\ref{yexp}), the coefficients
$D^{(1)}_{[i]}$ are explicitly
\eq
D^{(1)}_{[i]}(E,{\bf g})=
(-1)^{i}\,\frac{ W^{(n)} [ W^{(1)}_0(x,E,{\bf g}), \chi_0,
\dots \chi_{i-1},\chi_{i+1} ,\dots \chi_{n-1}]}
{ W^{(n)}[\chi_0,\dots \chi_{n-1}]}
\label{di}
\en
where 
\eq W^{(n)}[\chi_0,\dots \chi_{n-1}]=
\prod_{j=0 , i=j+1 }^{n-1} (g_j - g_i)~.
\en
Thus, for example, continuation of the parameters from
${\bf g}=(g_0,\dots g_i,\dots g_j,\dots g_{n-1})$ to 
${\bf  g'}=(g_0,\dots g_j, \dots g_i, \dots g_{n-1})$
has the following effect
\eq
D^{(1)}_{[i]}(E,{\bf    g'}) =D^{(1)}_{[j]}(E,{\bf    g})
\ ,\qquad D^{(1)}_{[j]}(E,{\bf  g'}) = D^{(1)}_{[i]}(E,{\bf    g})\,,
\en
with the other $D^{(1)}_{[k]}(E,{\bf   g})$ remaining unchanged.
So long as the property (\ref{van}) is preserved at all points of the 
continuation,
the NLIE will hold throughout. At the end of the continuation, the ODE 
will have
returned to its original form, but the NLIE will have undergone a nontrivial
monodromy, since the twists will have changed.
Explicitly, an NLIE describing 
$D^{(1)}_{[i]}(E,{\bf g})$ can be found by continuing the twist parameter
${\bf \alpha}$ to ${\bf  \alpha'}$
via the transformation $g_0 \to g_i$, with the remaining $g_k$
being allowed to permute amongst themselves.
In fact, a number of different NLIEs can be obtained, 
depending on the permutation chosen. 

A set of special choices for
the vector ${\bf g}$ provides a simple illustration of all this.
We use the notation 
${\bf   \hat g}$ to indicate a vector with distinct components $g_j$
taking values in the set $\{0,1,\dots  (n-1)\}$. 
 There are
$n!$ different ${\bf \hat  g}$, but for each choice the resulting
differential operator 
$D({\bf \hat g})$ is equal to $\frac{d^n }{dx^n}\,$.  We
will use ${\bf \hat g}_i$  to indicate $g_0$ has been fixed to
be equal to $i$.
In this particularly
simple case,
the functions (\ref{di}) become 
\eq
D^{(1)}_{[i]}(E,{\bf \hat  g}_i)=\frac{1}{i!}
\frac{d^{i}} {dx^{i}}y(0,E,{\bf  \hat g})~.
\en
(Since the function $y(x,E,{\bf  \hat g})$ does not depend on the ordering
of the components 
of ${\bf  \hat g}$, we omit the subscript $i$ on the RHS of this equation.)
The direct treatment of previous sections only allowed us to
discuss $D^{(1)}_{[0]}=y(0,E,{\bf\hat g})$, but
according to the continuation idea just described, 
the NLIEs should also encode information about the higher derivatives
$y^{[i]}$.

To make a numerical check, we set $M=1/n$, so that $P(x,E)=x-E$ and the ODE
is solvable by a complex Fourier transform, as in \cite{DTc} for $n=3$.
This lies outside the range $M>1/(n{-}1)$, but,  
just as in previous examples \cite{DTa,DTc}, 
we assume that the final NLIE continues to hold. If we write
$y(x,E,{\bf \hat g}_i)= \CA(x{-}E) $, then the 
ODE becomes
\eq
(-1)^{n+1} \frac{d^n\CA(x)}{dx^n}+x\CA(x)=0
\en
and has the normalised solution
\eq
\CA(x)={1 \over i^{(n{-}1)/2}\sqrt{2\pi} }
 \int_{\Gamma} dp \
   e^{-ipx +\frac{(ip)^{n+1}}{n+1}}~.
\label{lin}
\en
The contour $\Gamma$ follows a path in the lower half  
complex plane that starts at $|x|=\infty$ along the ray
$\arg(x) = -\frac{\pi}{2} -\frac{\pi}{n{+}1}$
and ends along
$\arg(x) = -\frac{\pi}{2} +\frac{\pi}{n{+}1}$
(for $n{=}3$ this is illustrated in figure 2 of \cite{DTc}). 
Evaluating (\ref{lin}) numerically for
$n=3,4 $ and $5$,
we achieved good agreement with results obtained via  
the numerical treatment of (\ref{nlie}), not only for the zeroes
of the functions $y(0,E,{\bf \hat g})$ but also, after analytic
continuation from ${\bf g}_0$ to ${\bf g}_1$, ${\bf g}_2$ and so on,
of $y'(0,E,{\bf \hat g}),\dots,
y^{[n-1]}(0,E,{\bf \hat g})$.
The accuracy obtained was typically to $12$ digits, and could doubtless be
improved with longer computing runs.
Occasionally, the NLIE (\ref{nlie}) breaks down, and in particular attempts to solve it by numerical iteration fail to converge.
Generally this happens whenever there is a breakdown in one or other
of the two assumptions necessary for the derivation of the NLIE in its
simplest form, mentioned at the beginning of \S6 above.
For the continuations between the different 
${\bf \hat g}_i$ it seems to be the second assumption which
sometimes fails, with
points appearing on the positive real
axis of the complex $E$ plane, not in the set $\{E^{(m)}_k\}$,
at which the functions $a^{(m)}$ take the value $-1$.
(For $n=2$
these points are zeroes of 
$T(E)$,  and the necessary modification of the NLIE
was discussed in \cite{Fran}.)
In most cases we found that by altering the choice of ${\bf \hat g}_i$
the problem could anyway be avoided. One exceptional case 
is $n{=}3$.
The two choices for ${\bf \hat g}_1$ are $(1,0,2)$ and $(1,2,0)\,$, 
and for neither of these did iteration of the basic NLIE converge.
Despite this fact, an
extrapolation allowed
a comparison with the results of the complex Fourier transform to be made,
and we found that the zeroes of $y'(0,E, {\bf  \hat g})$ were indeed 
reproduced.
Note that in this case the zeroes of $y(0,E,{\bf  \hat g})$ and 
$y''(0,E, {\bf
\hat g})$ can be found from an NLIE which respects the $SU(3)$ diagram
symmetry~\cite{DTc}, but that to find the zeroes of 
$y'(0,E, {\bf \hat g})$, we
are forced to use an integral equation which breaks this symmetry.

\smallskip
\resection{Duality}
As in \cite{BLZa,DTb,DTc}, a duality 
transformation can be defined which relates solutions of
the ODEs with $M>0$
with solutions of ODEs of the same form but with $-1<M<0$.
For $n=2$, the case of the Schr\"odinger
equation, this
relates a confining potential to a 
singular potential.  The first step is to implement
a (generalised)
Langer \cite{La} transformation 
\eq
z=\ln x
\,,\quad 
y(x)=e^{(n-1)z/2}u(z)~.
\en
The use of the operators $D(g)$ and $D({\bf g})$ makes this task 
very simple, since the relation 
\eq
\left(\frac{d}{dx}-\frac{g}{x}\right) e^{\frac{p-1}{2}z(x)} f(z(x))=
e^{\frac{p-3}{2}z}\left(\frac{d}{dz}+\frac{p{-}1}{2}-g\right) f(z)
\en
allows exponentials of $z$
to be passed successively through the
factors of $D({\bf g})$, resulting in the transformed equation
\eq
\Bigl[(-1)^{n+1} \Bigl(\frac{d}{dz}-\g_{n-1}\Bigr)\dots
\Bigl(\frac{d}{dz}-\g_{1}\Bigr)
\Bigl(\frac{d}{dz}-\g_{0}\Bigr)
+e^{n(1+M)z}-Ee^{nz}\Bigr]u(z)=0~
\en
where $\g_i=g_i-(n{-}1)/2$.
The exponentials
can now be exchanged
via 
$ z  \to \frac{z}{M+1} + \ln\frac{M+1}{E^{1/n}}$ to obtain
\eq
\Bigl[ (-1)^{n+1}
\Bigl(\frac{d}{dz}-\frac{\g_{n-1}}{M{+}1}\Bigr)\dots
\Bigl(\frac{d}{dz}-\frac{\g_{1}}{M{+}1}\Bigr)
\Bigl(\frac{d}{dz}-\frac{\g_{0}}{M{+}1}\Bigr)
-\wt E e^{nz}-e^{\frac{nz}{M+1}}\Bigr]u(z)=0~, 
\en
where $\tilde{E}=-(M{+}1)^{nM}E^{-(M+1)}\,$.
Applying the inverse transformation,
(\ref{gnde}) becomes
\eq
\Bigl((-1)^{n+1} D({\bf \wt g})  
+P(x,\wt E) \Bigr)y(x)=0~,\qquad 
P(x,\wt E)= -x^{n\wt M}-\wt E~,
\en
with
\eq
\wt M=-\frac{M}{M{+}1}~,\qquad
{\bf \wt g}=\{\wt g_0, \wt g_{1} ,\dots ,\wt g_{n-1}\}~,\qquad
\wt g_i = \frac{g_i+M(n{-}1)/2}{M+1}~.
\en
(Observe that the property (\ref{van}) is preserved by the transformation
of ${\bf g}$.)
This, the dual equation, is the same as the original ODE (\ref{gnde}) apart 
from the reversed sign in front of the `potential' term
$x^{n\wt M}$, and the fact that the
parameters $(M,{\bf g})$ have been replaced by
$(\wt M,{\bf\wt g})$.
Thus the equation
transforms very simply
under duality.
\smallskip
\resection{Conclusions}
In this paper we have found a hidden $SU(n)$ structure
inside certain $n^{\rm th}$-order differential 
equations. This has enabled a large set of Bethe Ansatz systems, and their
associated nonlinear integral equations, to be related directly to
spectral problems. 

One aspect that we did not discuss relates to the truncations of the 
functional
relations which can occur at special values of the parameters $M$ and 
${\bf g}$. 
These can provide a link to equations arising in the thermodynamic Bethe 
ansatz,
a subject of much independent interest. The earlier analysis of \cite{Sb}, 
which
should be considered as complementary to our work, was much closer to this
approach, and it would be interesting to develop it further.

The correspondence between Bethe ansatz systems and ordinary 
differential equations
has various potential applications.
{}From the point of view of integrable systems, results such as
duality properties are made much more apparent by 
the mapping into a differential equation, a fact which has already been
exploited, for $n{=}2$, in~\cite{BLZa}.
In fact,
at least for $n=2$, 
the $T$ and $Q$-functions are coming to have an increasingly
 important role to play in the general study
of integrable quantum field theories with  
boundaries~\cite{FLS1,BLZ1,FLS2,DAPTW,LSS,DRTW}, and this makes it likely that
the new perspective on these functions provided by their reinterpretation as
spectral determinants will prove of broader relevance.
(This subject is even
starting to find applications in string theory \cite{HKM}.) 
To give one example, in the study of boundary flows
direct relations between the $T$'s and
the $g$-functions of Affleck and Ludwig~\cite{AL}
have been found in certain cases \cite{FLS1,BLZ1,DRTW}.

{}On the other hand, it is also
rather remarkable that one can obtain information about spectral determinants
for quantum {\em mechanical} problems, and more general ODEs, using 
conformal field theory techniques such as  the truncated conformal
space approach~\cite{DAPTW} and  perturbation theory~\cite{BLZa,BLZ1,DRTW}.
Our treatment also entailed some rather strong
conjectures about the reality and positivity 
 properties of spectral problems associated 
with the ordinary differential equations under consideration.
These conjectures were
well-supported by
numerical work, but proofs are still lacking.
Previous examples of spectral problems which, despite having no obvious
hermiticity, still exhibit real spectra include the ${\cal CP}$ symmetric 
quantum
mechanics of~\cite{BB}. (This was briefly
discussed from the point of view of functional
relations and T-Q systems in~\cite{DTb}.) It appears that a much larger set of
problems sharing similar properties is being uncovered, and it would be very
rewarding to understand this behaviour at a deeper level.

%
%
%
\medskip
\noindent{\bf Acknowledgements --} 
We would like to thank John Cardy, David dei Cont,
Philippe di Francesco, Jan de Gier, Bernard Nienhuis,
Andr\'e Voros,  Ole Warnaar and Jean-Bernard Zuber 
for useful conversations and help. 
PED and TCD thank SPhT Saclay,
and PED thanks the Departamento de F\'\i sica
de Part\'\i culas,
USC (Santiago de Compostela), for hospitality.
RT thanks the Universiteit van Amsterdam for a post-doctoral fellowship,
and PED and TCD thank the EPSRC for an Advanced Fellowship and a
Research Studentship respectively.
The work was supported in part by a TMR grant of the
European Commission, reference ERBFMRXCT960012.

\medskip
\noindent{\bf Note added --} After this paper was finished 
we were informed by
J.~Suzuki that he has also been  working on this topic, with results
which partially overlap those reported here.
We also thank J.~Suzuki for suggesting some improvements in our
terminology.
%

\end{document}